\documentstyle[epsfig,longtable]{aipproc}
\begin{document}
\title{Black Hole Emergence in Supernovae}

\author{Shmuel Balberg$^1$, Stuart L.~Shapiro $^{1,2}$ 
and Luca Zampieri$^{1,3}$}
\address{
$^1$Department of Physics, University of Illinois at 
Urbana--Champaign, Urbana, IL \\
$^2$Department of Astronomy and National Center for
                                         Supercomputing Applications 
             University of Illinois at Urbana--Champaign, Urbana, IL\\
$^3$Department of Physics, University of Padova, Padova, Italy}
\maketitle
\vspace{-0.5cm}
\begin{abstract}
If a black hole formed in a core-collapse supernova is accreting material 
from the base of the envelope, the accretion luminosity could be 
observable in the supernova light curve. We present results of  
a fully relativistic numerical investigation of the 
fallback of matter onto a black hole in a supernova and examine conditions 
which would be favorable for detection of the black hole. 
In general, heating by radioactive decays is likely to prevent 
practical detection of the black hole, but we show that low energy explosions
of more massive stars may provide an important exception.  
We emphasize the particular case of SN1997D in NGC1536, 
for which we predict that the presence of a black hole could be inferred 
observationally within the next year.   
\end{abstract}
\vspace{-0.2cm}
\section*{Introduction}
%
%
%
%
Theory suggests that the compact object formed in a core-collapse supernova 
can be either a neutron star or a black hole, depending on the character 
of the progenitor and the details of the explosion \cite{Fryer-here}.    
The presence of several radio pulsars in sites of known supernovae provides 
substantial observational evidence that neutron stars are indeed created 
in supernovae, but similar evidence for a black hole - supernova connection 
is still mostly unavailable (see \cite{Israelian99} for recent 
indirect evidence).

A newly formed black hole in a supernova can be identified directly if it 
imposes an observable effect on the continuous emission of light that follows 
the explosion - the {\it light curve}. 
In particular, if some material from 
the bottom of the expanding envelope remains gravitationally bound to the 
black hole, it will gradually fall back onto it, generating an accretion
luminosity \cite{ZCSW98}.  
The black hole can be said to ``emerge'' in the supernova light 
curve if and when this luminosity becomes comparable 
to the other sources that power the light curve. 


\section*{Black Hole Emergence in the Light Curve}

Since the material which remains bound to the black hole following 
a supernova is outflowing in an overall expansion, the accretion rate 
must decrease in time. The expansion will also cause pressure forces 
to become unimportant eventually, and 
the accretion will proceed as dust-like, following a power-law decline in 
time according to $\dot{M}\propto t^{-5/3}$ \cite{CSW96}.
As shown in \cite{ZCSW98}, the accretion flow and the radiation field
proceed as a sequence quasi-steady-states, and  
the accretion luminosity can therefore be 
estimated according to the formula of Blondin \cite{Blondin86} for 
stationary, spherical, hypercritical accretion onto a black hole
($L\propto \dot{M}^{5/6}$). The accretion luminosity then takes the form
\cite{ZCSW98,ZSC98,BSZ99}:
\begin{equation}\label{eq:L_acc}
L_{acc}(t) \propto 
L_{acc,0}t^{-25/18}\;,
\end{equation}
where $L_{acc,0}$ depends on the kinetic energy, density and composition 
of the accreting material at the onset of dust-like flow.

Heating by decays of radioactive elements synthesized in the explosion 
may provide a significant source of luminosity in the late-time light curve. 
The time dependence of radioactive heating rate for an isotope $X$ may 
be estimated as \cite{Woosleyal89}
\begin{equation}\label{eq:Q_rad}
Q_X(t)=M_X \varepsilon_X f_{X,\gamma}(t) \mbox{e}^{-t/\tau_X}\;, 
\end{equation}
where $M_X$ is the total mass of the isotope $X$ in the envelope, $\tau_X$ is 
the isotope's life time, and $\varepsilon_X$ is 
the initial energy generation rate per unit mass.
The factor $f_{X,\gamma}(t)$ reflects that not all $\gamma-$rays emitted in 
the decays are efficiently trapped in the envelope (and so do not 
contribute to the UVOIR luminosity).

Since accretion luminosity decreases as a power law in time 
while radioactive heating declines exponentially, then - assuming 
that spherical accretion persists - the accretion luminosity must eventually 
become the dominant source in the light curve. Furthermore, 
the non-exponential character of the accretion luminosity should be readily 
distinguishable in observations, announcing that the black hole has 
``emerged'' in the light curve.

\section*{Realistic Supernovae}

The typical amount of radioactive elements observed in type II supernovae 
suggests that an observation of black hole 
emergence in the light curve will usually be impractical. 
For example, luminosity due to accretion onto 
a hypothetical black hole in SN1987A
would become comparable to the heating rate due to 
positron emission in $^{44}$Ti decays only $\sim\!900\;$years after the 
explosion. At this time the luminosity will have dropped to only 
$\sim10^{32}\;\mbox{ergs}\;\mbox{s}^{-1}$ \cite{ZCSW98}.

An important exception is expected in the case of higher mass progenitors, 
$M_*=25-40\;M_\odot$. Explosions of such stars are likely to involve 
significant early fallback even while the explosion is still proceeding. 
The survey of Woosley and Weaver 
\cite{WoosWeav95} suggests that, in general, larger mass stars leave 
behind larger remnants and expel a smaller amount of radioactive isotopes
(since these are synthesized in the deepest layers of 
the envelope, and a significant fraction is advected back onto the collapsed 
core). Clearly, for such an explosion, there 
is likely to be a larger reservoir of 
bound material for late time accretion, so that combined with the low 
background of radioactive isotopes, an actual detection of black hole 
emergence may become feasible. We have recently conducted a numerical 
investigation of the expected emergence of a black hole in such supernovae
\cite{BSZ99}. This investigation was carried out with the 
spherical, fully relativistic 
radiation-hydrodynamics code described in \cite{ZCSW98}, modified to include 
a variable chemical composition with a detailed photon opacity table, and 
to account for radioactive heating.

\subsection*{Black Hole Emergence in ``Radioactive-Free'' Supernovae}

The most favorable case for  identifying black hole emergence in supernova
would be a low-to-medium energy ($\leq1.3\times 10^{51}\;$ergs) explosion 
of a progenitor star with a mass of $35-40\;M_\odot$, where the ejected 
envelope is expected to be practically free of radioactive isotopes 
\cite{WoosWeav95}. 
For such supernovae, the black 
hole should emerge within a few tens of days after the explosion. 
As an example, we show in Fig.~\ref{fig:lites} the calculated light curve of 
such an explosion, based on the theoretical model S35A of 
\cite{WoosWeav95} ($M_*\!=\!35\;M_\odot\;,\;M_{BH}\!=\!7.5\;M_\odot$). The 
luminosity at emergence is $\gtrsim 10^{37}\;\mbox{ergs}\;\mbox{s}^{-1}$, 
after which the 
light curve clearly follows a power law decline in time. If such a supernova 
were observed, it would offer an explicit opportunity to confirm the 
presence of a newly formed black hole \cite{ZCSW98}.

\subsection*{SN1997D}

While such an ideal candidate 
is not available at present, 
SN1997D may provide a marginally {\it observable} case for idenfying the 
emergence of a black hole.
Discovered on January 14, 1997 
in the  galaxy NGC 1536,  
SN1997D is the most sub-luminous 
type II supernova ever recorded. Through an analysis of the light curve and 
spectra, Turatto et al.~\cite{Turatto98} suggested that the supernova was 
a low energy explosion, $\sim 4\times 10^{50}\;$ergs, of a 
$26\;M_\odot$ 
star. 
The observed 
late-time light curve (up to 416 days after the explosion) is consistent 
with only $\sim\!0.002\;M_\odot$ of $^{56}$Co in the ejected envelope, 
much lower than the $\sim0.1\;M_\odot$ typical of most type II supernova 
($0.075\;M_\odot$ in SN1987A). In a 
preliminary investigation, Zampieri et al.~\cite{ZSC98} pointed out that the 
$3\;M_\odot$ black hole (prdecited for this model) 
may emerge in SN1997D as early as $\sim 3\;$years after 
the explosion, with an accretion luminosity ranging between 
$10^{35}$ to as much as $5\!\times\!10^{36}\;\mbox{ergs}\;\mbox{s}^{-1}$.

\begin{figure}[t]
\begin{minipage}[t]{68mm}
\begin{center}
\caption{
Light curves including accretion lumninosity for a $35\;M_\odot$ progenitor 
(model S35A of [9]), the best fit model of [10] 
for SN1997D (model I), and a variant of SN1997D where the initial conditions 
allow for a larger late time luminosity (model Ia).}
\label{fig:lites}
\includegraphics[angle=270,width=6.8cm]{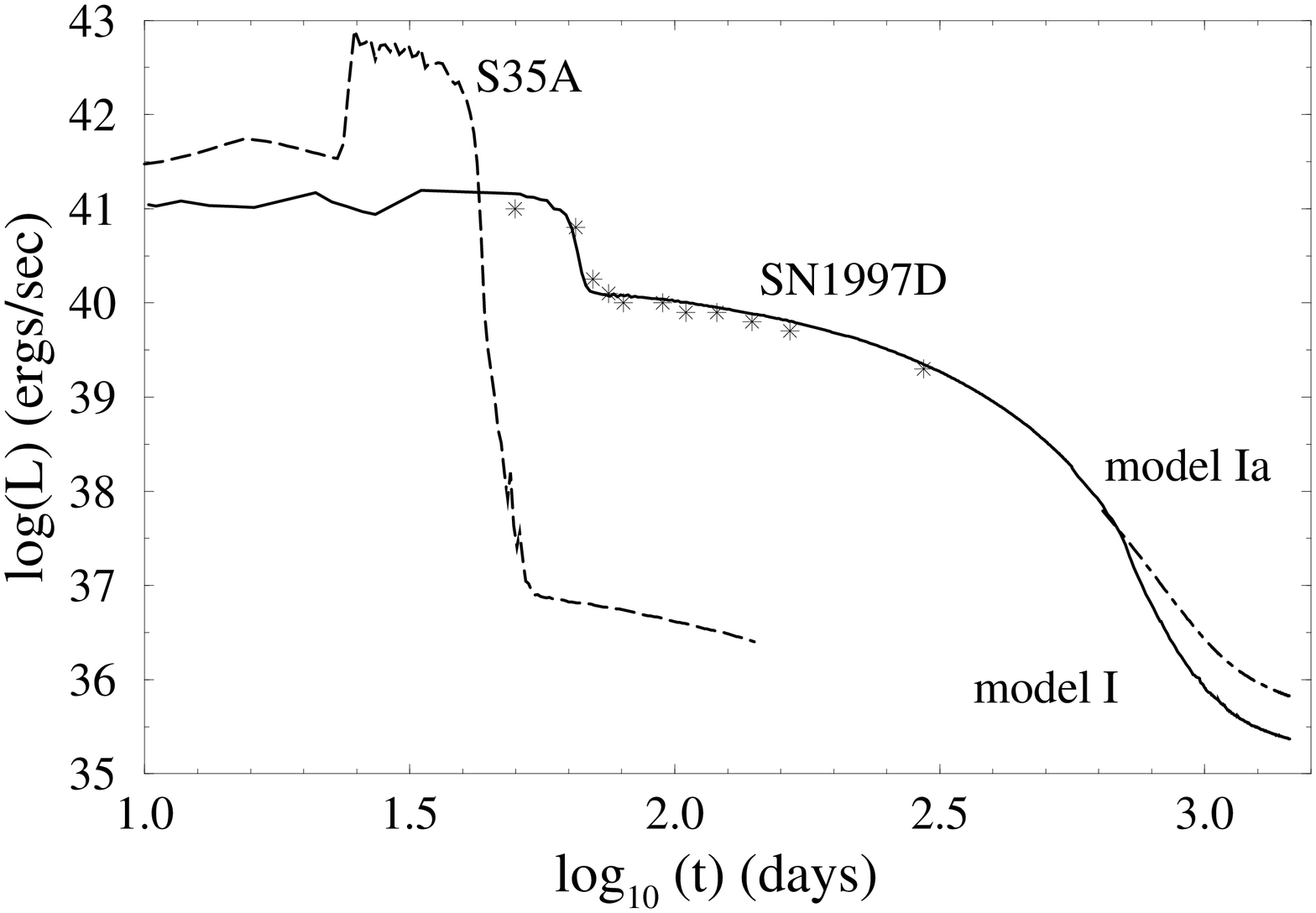}
\end{center}
\end{minipage}
\hspace{1cm}
\begin{minipage}[t]{68mm}
\begin{center}
\caption{
The total luminosity of model I for SN1997D and the contributions to 
the light curve by radioactive heating of $^{56}$Co, 
$^{57}$Co and $^{44}$Ti, their total ($L_{rad-tot}$), and the accretion 
luminosity. The arrow marks the time at which $L_{acc}=\frac{1}{2}L_{tot}$.
\label{fig:endgame}}
\vspace{-0.5cm}
\includegraphics[angle=270,width=6.8cm]{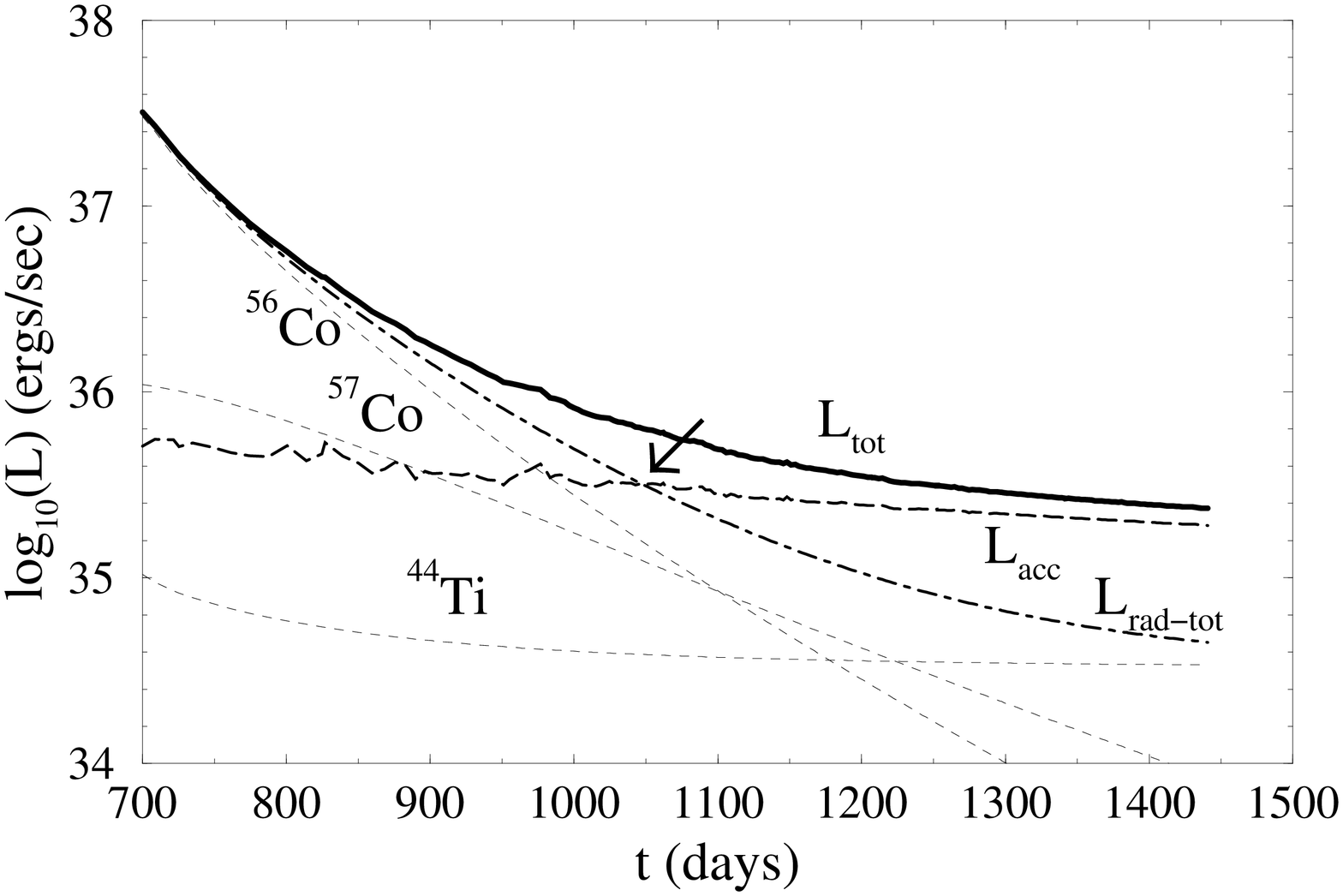}
\end{center}
\end{minipage}
\end{figure}

Our calculated light curve for SN1997D based on the best-fit post-explosion 
model of \cite{Turatto98} is shown in Fig.~\ref{fig:lites}.  
The earlier part of the light curve is in good agreement with 
the observed data, while at a later time we find that the black hole emerges 
about 1050 days after the explosion 
- which corresponds to late 1999 - {\it NOW!}. 
Figure \ref{fig:endgame} compares the 
heating due to the isotopes $^{56}$Co, $^{57}$Co and $^{44}$Ti 
to the accretion luminosity.
Note that
radioactive heating (especially $^{44}$Ti) is 
never negligible with respect to the accretion luminosity, so the total 
luminosity does not fall off as 
an exact power law. Nonetheless, the 
presence of the black hole could still be inferred by attempting to 
decompose the total light curve.

In this calculation, the total luminosity at emergence is about 
$7\!\times\!10^{35}\mbox{ergs}\;\mbox{s}^{-1}$. However, this luminosity is 
dependent on the finer 
details of the initial profile, which are difficult to constrain from the 
early light curve. 
Considering these uncertainties and those regarding the 
abundances of $^{57}$Co and $^{44}$Ti, 
we find through a revised analytical estimates 
(which account for $\gamma$-ray transparency and Eddington-rate limits on 
the accretion flow), that while the time of emergence is fairly 
well determined, the 
plausible range for the luminosity at emergence is 
$0.5-2\!\times\! 10^{36}\;\mbox{ergs}\;\mbox{s}^{-1}$. One example, where the 
luminosity at emergence is 
$\sim\!1.4\!\times\! 10^{36}\;\mbox{ergs}\;\mbox{s}^{-1}$ is also shown 
in Fig.~\ref{fig:lites}. 
In this case, the accretion luminosity is sufficiently high that
contribution of radioactive heating does not cause any significant
deviation from a power-law decay.
We estimate that such a luminosity is still 
marginally detectable ($m_v\!\approx\!29$) with the HST STIS camera, 
so that if observed, SN1997D could provide 
first direct observational evidence of 
black hole formation in supernova within the next year.

\end{document}